\documentclass[a4paper,fleqn]{article} 
\usepackage{modsim}
\usepackage{times}
\usepackage{natbib} 
\usepackage{amsmath, amssymb, amsthm} 

\MODSIMhead{P.V. Shevchenko, J. Hirz and U. Schmock, Forecasting Leading Death Causes using Extended CreditRisk$+$} 

\usepackage{rotating}
\usepackage{amsbsy,enumerate}
\usepackage{graphicx}
\usepackage{ccaption}

\makeatletter
\renewcommand{\fnum@figure}[1]{\textbf{\figurename~\thefigure}. }
\renewcommand{\fnum@table}[1]{\textbf{\tablename~\thetable}. }
\makeatother

\begin{document}

\title{Forecasting Leading Death Causes in Australia using Extended CreditRisk$+$}

\author{\underline{P.V. Shevchenko} \address[A1]{\it CSIRO Risk Analytics Group, Australia}, J. Hirz \address[A2]{\it{Department of Financial and Actuarial Mathematics, Vienna TU, Austria}} and U. Schmock \addressmark[A2]\\\textnormal{ }\\{\textnormal{26 July 2015}}} 

\email{Pavel.Shevchenko@csiro.au} 

\date{26 July 2015}


\begin{abstract}
	Recently we developed a new framework in \citet{HirzSchmockShevchenko} to model stochastic mortality using extended CreditRisk$^+$ methodology which is very different from traditional time series methods used for mortality modelling previously. In this framework, deaths are driven by common latent stochastic risk factors which may be interpreted as death causes like neoplasms, circulatory diseases or idiosyncratic components. These common factors introduce dependence between policyholders in the annuity portfolios or between death events in population. This framework can be used to construct life tables based on mortality rate forecast. It also provides an efficient, numerically stable algorithm for an exact calculation of the one-period loss distribution of annuities or life insurance products portfolios and associated risk measures such as value-at-risk and expected shortfall required by many regulators. Moreover this framework allows stress testing and, therefore, offers insight into how certain health scenarios influence annuity payments of an insurer. Such scenarios may include improvement in health treatments or better medication.
In this paper, using publicly available data for Australia, we estimate the model using Markov chain Monte Carlo method to identify leading death causes across all age groups including long term forecast for 2031 and 2051. On top of general reduced mortality, the proportion of deaths for certain certain
causes has changed massively over the period 1987 to 2011. Our model
forecasts suggest that if these trends persist, then the future
gives a whole new picture of mortality for people aged above 40 years.
Neoplasms will become the overall number-one death cause. Moreover, deaths due to mental and behavioural disorders are very likely to surge whilst deaths due to
circulatory diseases will tend to decrease. This potential increase in deaths due to mental
and behavioural disorders for older ages will have a massive impact on social systems
as, typically, such patients need long-term geriatric care.
\end{abstract} 

\maketitle

\section{INTRODUCTION}
Mortality modelling has a very long history. Numerous deterministic survival models for mortality intensity have been developed since Benjamin Gompertz suggested an exponential increase in death rates with age (\emph{Gompertz' law of mortality}) in 1825. However, stochastic modelling of mortality is more modern development over last twenty years. One of the standard benchmarks a model developed by \citet{Lee-Carter}, for modelling log death rates of people across different ages affected by one time dependent latent factor. There are many extensions introducing several factors and special treatment for cohort effects; for review, see \citet{booth2008mortality}.
This stochastic modelling has become increasingly important in the financial industry and government departments since it is observed over the last several decades that life expectancy is typically underestimated. Moreover, new regulatory requirements such as Basel III and Solvency II and the financial crises of 2008 also call for stochastic mortality modelling.

\begin{figure}[htp]
		\begin{center}
		\includegraphics[width=0.8\textwidth]{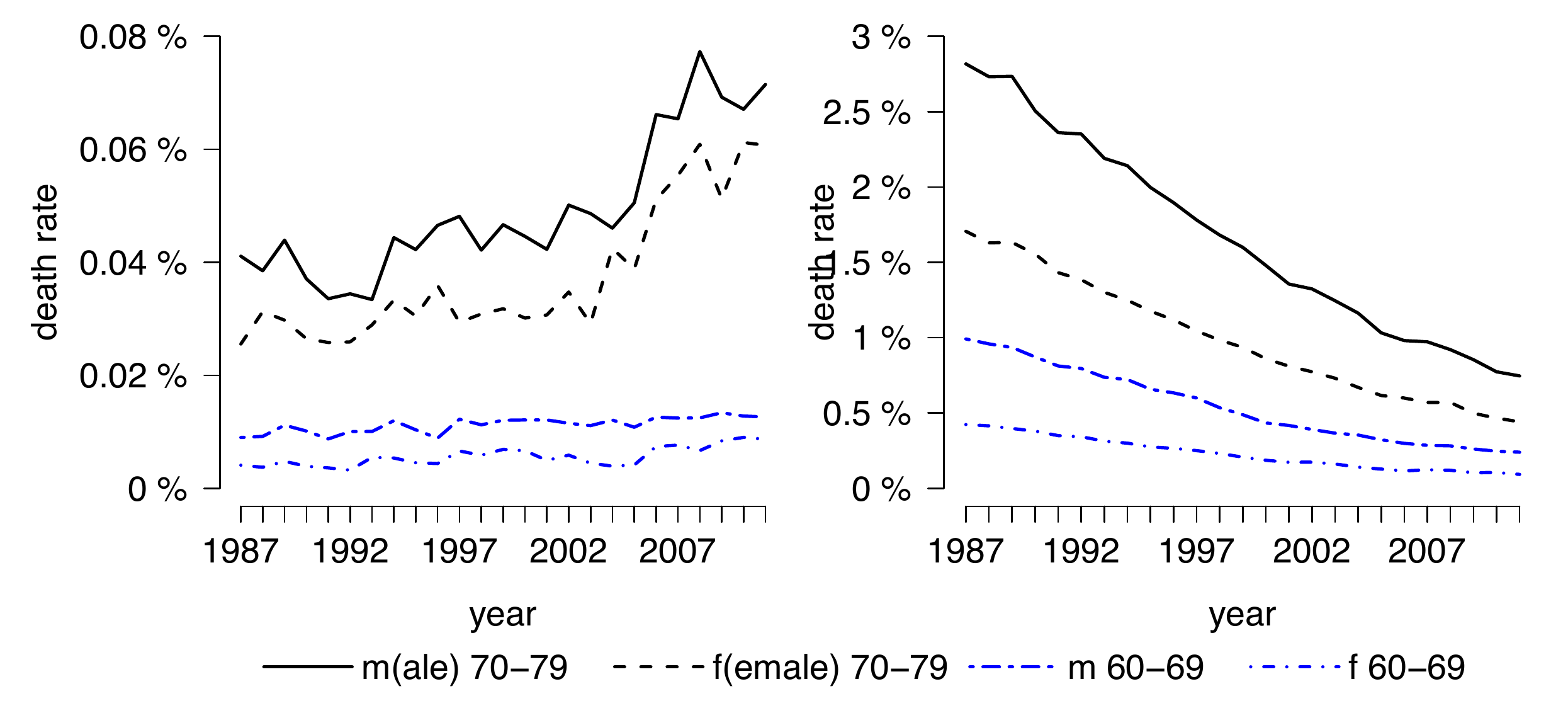}
		\caption[Australian death rates for mental and behavioural disorders 1987--2011]{Australian death rates for mental and behavioural disorders (left), as well as for
		circulatory diseases (right) from 1987 to 2011 for
		age categories\/ 70--79 years, and 60--69 years with both genders.}
		\label{fig:death_causes}
		 \end{center}
	\end{figure}

Life insurers and pension funds usually use deterministic first-order life tables to derive premiums, forecasts, risk measures for portfolios and other related
	quantities. These first-order life tables are derived from second-order life tables (best estimates of the current mortality of a population) plus artificially added risk margins associated with longevity, size of the company, selection phenomena, estimation and various other sources, see, for example, \citet{DAV}.
	The risk margins described there
	 often lack stochastic foundation and are certainly not consistently appropriate
	for all companies due to a possibly twisted mix of these risks. Moreover,
    drastic shifts in death rates due to certain death causes are observed over the past decades. This phenomenon is usually
	not captured by generation life tables which incorporate only an overall trend in death probabilities. As an
	illustration of this fact, Figure \ref{fig:death_causes} shows death rates based on Australian data.

In our recent paper \citet{HirzSchmockShevchenko}, a collective risk model called extended CreditRisk$^+$ was adapted to modelling mortality and calculation of loss distribution for annuity portfolios. This model setup is very different from traditional time series models developed for mortality modelling previously.
As the name suggests, it is a credit risk model used to derive loss distributions of credit portfolios
	and originates from the classical CreditRisk$^+$ model which was introduced by Credit Suisse First Boston (\citeyear{CRP}). Within credit risk models it is classified as a Poisson mixture model. Identifying default with death makes the model perfectly applicable for various
	kinds of life insurance portfolios and annuity portfolios.

In this paper, using publicly available data for Australia, we estimate model parameters via a Markov chain Monte Carlo (MCMC) method to identify leading death causes across all age groups including long term forecast for 2031 and 2051. Section \ref{sec:Model} presents a model formulation. Likelihood function and  MCMC method for estimation  are summarised in Section \ref{sec:Estimation}. Results are presented in Section \ref{sec:RealData} and Section \ref{sec:Conclusion} contains concluding remarks.

\section{Model}\label{sec:Model}
We follow the model developed in \citet{HirzSchmockShevchenko} where the reader can find precise mathematical formulation. Here we provide a brief outline of the main model blocks and assumptions.

Let $\{1,\dots,m\}$ denote	the set of people in population (or policyholders in the annuity portfolio)
	with death indicators $N_1,\dots,N_m$ over some period of time (i.e. one year if we model annual death rates or annual portfolio loss distribution).
	Event $N_i=0$ indicates no death for $i\in\{1,\dots,m\}$. In reality, death indicators $N_i$ are Bernoulli variables as each person can die only once. However for easier calibration and subsequent calculations a Poisson approximation is assumed  that does not lead to material difference in the case of large number of people in a group as death probabilities over one year are small, see discussions and numerical examples in \citet{HirzSchmockShevchenko}.

Denote latent stochastic risk factors identified with causes of death (such as neoplasms, cardiovascular diseases or idiosyncratic components) by $\Lambda_k$, $k=1,\ldots,K$ which are modelled as independent random variables from gamma distribution with mean 1 and variances $\sigma_k^2$, $k=1,\ldots,K$.
These stochastic risk factors are designed to model	effects which simultaneously
	influence death probabilities of many people due to a common exposure to the same type of risk. Risk index $k=0$ is reserved to represent idiosyncratic risk.
	
Denote corresponding weights (vulnerability of policyholder $i$ to risk
factor $k$) as $w_{i,k}$ and by model construction	$w_{i,0}+\dots +w_{i,K}=1$. In average, when we consider homogeneous groups of people this weight is the fraction of people dying from death cause $k$ compared to all deaths in the group.

For every person $i\in\{1,\dots,m\}$,
			the total number of deaths $N_i$ is split up additively according
			to risk factors as
\begin{equation}
				N_i=N_{i,0}+\dots+N_{i,K},
\end{equation}
i.e. $N_{i,k}$ is the number of deaths of a person $i$ due to risk factor $k$.
It is assumed that death indicators for idiosyncratic risk $N_{1,0},\dots,N_{m,0}$ are independent
from one another, as well as all other	random variables and $N_{i,0}$ is Poisson distributed with
				intensity $q_i w_{i,0}$, where $q_i$ denotes the probability of death of a person $i$.

Conditionally (given risk factors), death indicators $N_{i,k}$ are independent and Poisson distributed with random intensity $q_{i} w_{i,k} \Lambda_{k}$, i.e.,
\begin{equation}
					\Pr\bigg(\bigcap_{i=1}^m\bigcap_{k=1}^K \{N_{i,k}=n_{i,k}\}\,
					\bigg|\,\Lambda_1,\dots,\Lambda_K\bigg)=
					\prod_{i=1}^m\prod_{k=1}^K e^{-q_{i} w_{i,k} \Lambda_{k}}
					\frac{(q_{i} w_{i,k} \Lambda_{k})^{n_{i,k}}}{n_{i,k}!}.
\end{equation}
Thus, by our model construction, $\mathrm{E}[N_i]=q_i (w_{i,0}+\dots+w_{i,K})=q_i$. Note that under this model, loss distribution for portfolio of annuities can be calculated exactly and efficiently using Panjer recursion algorithm instead of approximate and slow Monte Carlo method; for more details, see \citet{HirzSchmockShevchenko}.

Consider time periods $t\in\{1,\dots,T,...\}$ where $T$ corresponds to the last year in the dataset.
To model trends in time, death probabilities $q_i$ and weights $w_{i,k}$ are assumed to be time dependent. If the modeller is interested in prediction over one year, then simple linear functions of time can be appropriate, however for long term forecasting these are designed more carefully as
\begin{equation}\label{eq:PDFamily}
q_{i}(t)=F^{\mathrm{Lap}}\big(\alpha_{i} +\beta_{i} \mathcal{T}_{\zeta_{i},\eta_{i}}(t)\big),\;\;w_{i,k}(t)= \frac{\exp\big(u_{i,k}+v_{i,k} \mathcal{T}_{\phi_{k},\psi_{k}}(t)\big)}{\sum_{j=0}^K \exp\big(u_{i,j}+v_{i,j}\mathcal{T}_{\phi_j,\psi_j}(t)\big)}\,,
\end{equation}
with model parameters $\alpha_{i},\beta_{i},\zeta_{i}\in\mathbb{R}$ and $\eta_{i}\in(0,\infty)$, as well as  $u_{i,0},v_{i,0},\phi_0,\dots,u_{i,K},v_{i,K},\phi_K\in\mathbb{R}$, and $\psi_{0},\dots,\psi_{K}\in(0,\infty)$  to be estimated using data. Here, $F^{\mathrm{Lap}}(x)$ is Laplace distribution and $\mathcal{T}_{\zeta,\eta}(t)$ is trend reduction function formally defined as
	\begin{equation}\label{LaplaceDistr}
		F^{\mathrm{Lap}}(x)=\frac{1}{2}+\frac{1}{2}\mathrm{sign}(x)\big(1-\exp(-|x|)\big)\,,\quad x\in\mathbb{R},
	\end{equation}
	\begin{equation}\label{CauchyDistr}
		\mathcal{T}_{\zeta,\eta}(t)=\frac{1}{\eta}\arctan(\zeta+\eta t)\,,\quad t\in\mathbb{R},\;(\zeta,\eta)\in\mathbb{R}\times(0,\infty).
	\end{equation}
For $x< 0$, Laplace distribution (\ref{LaplaceDistr}) becomes $\exp(x)/2$ and thus corresponds to modelling of log death rates. Trend reduction function is motivated by  \citet[section 4.6.2]{AVOE_annuity_table} where linear time trend was replaced by time shift $\mathcal{T}_{0,\eta}(t)$
		with $\eta=\frac{1}{t_0}$. 	Then, parameter $\eta$ gives the inverse of the time $t_0$ when an initial trend is halved.
		Parameter $\zeta$ on the other hand gives the shift on the arctangent curve.
Note that $\lim_{x\to\pm\infty}\arctan(x)=\pm\frac{\pi}{2}$ and thus death probabilities and weights (\ref{eq:PDFamily}) are non-degenerate, i.e., do not hit zero or one, as $t\to\infty$.

\section{ Data Likelihood and MCMC estimation}\label{sec:Estimation}
For every age category\/ $a\in\{1,\dots,A\}$, gender\/		
	$g\in\{\mathrm{f},\mathrm{m}\}$ and year\/ $t\in\{1,\dots,T\}$ with\/ $T\geq 2$ the database is assumed to contain historical population counts\/ $m_{a,g}(t)$ and
	historical number of deaths\/ $n_{a,g,k}(t)$
due to underlying death cause\/ $k\in\{0,1,\dots,K\}$. An underlying death cause is to be understood as the disease or injury
that initiated the train of morbid events leading directly to death.

 The observations of historical annual deaths
	$n_{a,g,k}(t)$ with age  $a\in\{1,\dots,A\}$, gender  		
	$g\in\{\mathrm{f},\mathrm{m}\}$, due to death cause  $k\in\{0,\dots, K\}$  and at time  $t\in\{1,\dots,T\}$ correspond to realisations of the random variable
	$$
		N_{a,g,k}(t):=\sum_{i\in M_{a,g}(t)}N_{i,k}(t)\,,
	$$
	where\/ $M_{a,g}(t)\subset\{1,\dots,m(t)\}$ denotes the set of  people (policyholders)
	of  specified age group and gender. Note that\/
	$N_{i,k}(t)$ is the number of deaths of a person  $i$ due to death cause  $k$ in year  $t$.
	Death cause  zero corresponds to ill-defined and not elsewhere reported deaths, i.e., idiosyncratic components, but could also be put as another stochastic risk factor.

It is assumed that for all $t\in\{1,\dots,T\}$, quantities $q_i(t)$ and corresponding weights $w_{i,k}(t)$, respectively, are the same for all	people $i\in\{1,\dots,m(t)\}$ within
			the same age category\/ $a$, same gender\/ $g$ and with respect to the same risk factor
			$\Lambda_k(t)$. Therefore, we can set	$q_{a,g}(t):=q_i(t)$ and
			$w_{a,g,k}(t):=w_{i,k}(t)$ for a representative policyholder\/ $i$ of age category\/ $a$ and gender\/ $g$ with respect to risk factor\/ $\Lambda_k(t)$.
	Define
	$$
		n_k(t):=\sum_{a=1}^A\sum_{g\in\{\mathrm{f},\mathrm{m}\}}n_{a,g,k}(t)\,,
	$$
	as well as\/ $\rho_{a,g,k}(t):=m_{a,g}(t) q_{a,g}(t) w_{a,g,k}(t)$ and
	$$
		\rho_k(t):=\sum_{a=1}^A\sum_{g\in\{\mathrm{f},\mathrm{m}\}}\rho_{a,g,k}(t)\,.
	$$
	Then, the likelihood function
 $\ell( \boldsymbol{n}|\boldsymbol{\theta}_q,\boldsymbol{\theta}_w,\boldsymbol{\sigma})$
	of parameters\/ $\boldsymbol{\theta}_q:=({\alpha},{\beta},\zeta,\eta)$, as well as
	$\boldsymbol{\theta}_w:=({u},{v},\phi,\psi)$ and $\boldsymbol{\sigma}:=(\sigma_1,\ldots,\sigma_K)$
	 given mortality data
	$ \boldsymbol{n}:=(n_{a,g,k}(t),a=1,\ldots,A,g\in\{\mathrm{f},\mathrm{m}\},t=1,\ldots,T)$
	can be easily calculated. Given risk factors $\Lambda_1,\ldots,\Lambda_K$ the conditional likelihood is just a product of appropriate Poisson probabilities, then the unconditional likelihood can be found by integrating out the gamma distributed independent $\Lambda_1,\ldots,\Lambda_K$ and is given by
	\begin{equation}\label{MLE_likelihood}
	\begin{split}
		\ell(\boldsymbol{n}|\boldsymbol{\theta}_q,\boldsymbol{\theta}_w,\boldsymbol{\sigma})=
		\prod_{t=1}^T \Bigg(&\bigg(\prod_{a=1}^A\prod_{g\in\{\mathrm{f},\mathrm{m}\}}\frac{e^{-\rho_{a,g,0}(t)} \rho_{a,g,0}(t)^{n_{a,g,0}(t)}}
		{n_{a,g,k}(t)!}\bigg)\\
		&\times		
		\prod_{k=1}^K\bigg(\frac{\Gamma(1/\sigma^2_k+n_k(t))}
		{\Gamma(1/\sigma^2_k) (\sigma^2_k)^{1/\sigma^2_k} (1/\sigma^2_k+\rho_k(t))^{1/\sigma^2_k+n_k(t)}}\\
		&\qquad\quad\,\,\times\prod_{a=1}^A\prod_{g\in\{\mathrm{f},\mathrm{m}\}}
		\frac{ \rho_{a,g,k}(t)^{n_{a,g,k}(t)}}
		{ n_{a,g,k}(t)!}\bigg)\Bigg)\,.
	\end{split}
	\end{equation}
It is also possible to derive closed form maximum a posteriori (MAP) estimates for risk factor realisations and their variances; for details of MAP estimates and likelihood derivation, see \citet{HirzSchmockShevchenko}.
	
Once the likelihood is calculated in closed form, it is straightforward to apply MCMC method to get many samples
${\boldsymbol\theta}^i=(\boldsymbol{\theta}^i_q,\boldsymbol{\theta}^i_w,{\boldsymbol{\sigma}}^i)$, $i=1,2,\ldots$ from the posterior density
$\pi({\boldsymbol\theta}|\boldsymbol{n})\propto \ell( \boldsymbol{n}|{\boldsymbol\theta})\pi({\boldsymbol\theta})$.
The mean over these samples provides good point estimate for the model parameters and sampled posterior density can be used to estimate parameter uncertainty. We use non-informative (uniform) prior density $\pi({\boldsymbol\theta})$ for model parameters; in this case the mode of the posterior samples corresponds to a maximum likelihood estimate and inference is based on data with non-material impact from the prior.
There are many MCMC algorithms that can be applied. In this study we implement a well known random walk Metropolis--Hastings within Gibbs algorithm as
described e.g. in \cite[section 7.4.4]{cruzpetersshevchenko}. Also, we utilize truncated normal distributions as proposal distributions.
The method requires a certain burn-in period until the generated chain becomes stationary. Estimates derived by matching of moments as described in \citet{HirzSchmockShevchenko}	can be used  as initial values to ensure a shorter burn-in period. To reduce long computational times, one can run several independent MCMC chains with different starting points on different CPUs in a parallel way.
We have tested our implementation via simulated experiments where we simulate risk factors and data from the model with known parameters and then estimate risk factors and parameters.

\section{Study of Australian mortality data}\label{sec:RealData}
We applied the above described model and MCMC estimation procedure to Australian death rate data for the period 1987 to 2011. Data source for historical Australian population, categorised by age and gender, is taken from the website of Australian Bureau of Statistic http://www.abs.gov.au and data for the number of deaths categorised by death cause and divided into nine age categories (0--9 years,
10--19 years, 20--29 years, 30--39 years, 40--49 years, 50--59 years, 60--69 years, 70--79 years and 80+ years, denoted by $a_1,\dots,a_9$, respectively) for each gender
is taken from the Australian Institute of Health and
Welfare website http://www.aihw.gov.au.

Data handling needs some care as there was a change in International Classification of Diseases in  Australia in 1997 as explained at the website of the ABS. As a result, for the period 1987 to 1996, death counts have to be multiplied by corresponding
comparability factors and rounded to the nearest integer in order to avoid
data inconsistencies. The provided death data is divided into 19 different death causes where we identify the following ten of them with common non-idiosyncratic risk factors (comparability factor is given in brackets): \emph{certain infectious and parasitic diseases} (1.25),
\emph{neoplasms} (1.0), \emph{endocrine, nutritional and metabolic diseases} (1.01), \emph{mental and behavioural disorders} (0.78),
\emph{diseases of the nervous system} (1.2), \emph{circulatory diseases} (1.0), \emph{diseases of the respiratory system} (0.91),
\emph{diseases of the digestive system} (1.05),
\emph{external causes of injury and poisoning} (1.06),
\emph{diseases of the genitourinary system} (1.14). We merge the remaining eight death causes to idiosyncratic risk, termed as \emph{not elsewhere defined}, as their individual contributions to overall death counts are small for all categories.

Trend reduction parameters are fixed a priori with values
$\zeta_{a_i,g}=\phi_k=0$, as well as $\eta_{a_i,g}=\psi_k=\frac{1}{150}$
for all ages $i\in\{1,\dots,9\}$, $g\in\{\mathrm{f},\mathrm{m}\}$ and
$k\in\{0,\dots,K\}$ with $K=10$.
Thus, we have to estimate $442$ parameters, $36$ of which can be chosen arbitrarily as the system is overdetermined. Fixing the parameter for trend reduction
makes estimation more stable and does not influence results significantly for mid-term
forecasts. The value of $\frac{1}{150}$ for $\eta_{a_i,g}$ and $\psi_k$ is an average approximation to trend reduction observed in Australia which is usually higher for very old and mid ages, see \citet{HirzSchmockShevchenko}.
Based on 35\,000 MCMC steps with burn-in period of 5\,000 we are able to derive estimates of all parameters.
 The results of estimation for leading death cause weightings including forecast for 2031 and 2051 across all age groups, using (\ref{eq:PDFamily}) and MCMC samples of the parameters in the usual way, are presented in Tables \ref{tab:LeadingCauses} and \ref{tab:LeadingCausesFem} for males and females, respectively. On top of general reduced mortality, the proportion of deaths for certain certain
causes has changed massively over the period 1987 to 2011. Our model
forecasts suggest that if these trends persist, then the future
gives a whole new picture of mortality for older ages. Weightings for death causes up to the age of $39$
are estimated to be relatively stable over time. Note that for the youngest age group, the leading death cause is idiosyncratic risk since it contains deaths related to the perinatal period, as well as
all congenital malformations, deformations and chromosomal abnormalities.
For ages above 40 years we derive an overall relative decrease in deaths due to circulatory diseases
with small estimation error (see MCMC quantiles in Tables \ref{tab:LeadingCauses} and \ref{tab:LeadingCausesFem}). Deaths due to neoplasms will become the overall number-one
death cause, but weights are relatively stable. Weightings for deaths due to mental and behavioural disorders show the most dramatic change and this cause will most likely become one of the leading death
cause for very old people. This potential increase in deaths due to mental
and behavioural disorders for older ages will have a massive impact on social systems
as, typically, such patients need long-term geriatric care. But it should be outlined
that estimation errors are high for this risk factor as can be seen in the wide confidence intervals
in Tables \ref{tab:LeadingCauses} and \ref{tab:LeadingCausesFem}.

As outlined in \citet{HirzSchmockShevchenko} using similar data, model validation techniques strongly suggest that our proposed model is suitable for
describing the observed data. These validation techniques include a test for checking whether sample covariances amongst deaths of the same cause lie within confidence intervals,
a test for independence amongst death counts of different death causes using a $t$-test, as well as tests for serial correlation and for distributional assumptions on risk factors.


\section{CONCLUSIONS}\label{sec:Conclusion}
This paper presents estimation results of leading death causes for all age groups in Australia including forecast for 2031 and 2051. The adopted modelling framework of extended CreditRisk$^+$ is very flexible. In a general form, extended CreditRisk$^+$ risk factors can be used to model risk
groups with simultaneous deaths of policyholders in the group (e.g. a couple dying in a car crash, people living near a volcano, virus
outbreaks), moreover dependence (negative and positive) between
death causes can be introduced. Development of this general case for mortality modelling is a subject of further research.

\section*{Acknowledgement}
P.~V.~Shevchenko gratefully acknowledges financial support by the CSIRO-Monash
Superannuation Research Cluster, a collaboration among CSIRO, Monash University, Griffith University, the University of Western Australia, the University of Warwick, and stakeholders of the retirement system in the interest of better outcomes for all.
J.~Hirz gratefully acknowledges financial support from the Australian Government via the 2014 Endeavour Research Fellowship, as well as from the Oesterreichische Nationalbank (Anniversary Fund, project number: 14977) and Arithmetica.

\begin{table}[!ht]
	\begin{center}\footnotesize{
		\caption{Leading death causes with weights estimated by MCMC sample average with 5 and 95 percent
		confidence intervals of MCMC posterior in brackets for \emph{males} of all age categories in 2011, 2031 and
2051.}
		\label{tab:LeadingCauses}
		\begin{tabular}{rr|rrr}
			\cline{3-5}\rule[-3pt]{0pt}{12pt}
			  && 2011 & 2031 & 2051  \\
			\hline\rule[10pt]{0pt}{0pt}
		&1.& not elsewhere: 0.669 $\big(\substack{0.679\\ 0.659}\big)$ &not elsewhere: 0.593 $\big(\substack{0.617\\ 0.559}\big)$& not elsewhere: 0.456 $\big(\substack{0.507\\ 0.378}\big)$\\[1ex]
		
		0--9 years& 2.& external: 0.108 $\big(\substack{0.114\\ 0.101}\big)$& circulatory: 0.077 $\big(\substack{0.103\\ 0.056}\big)$& circulatory: 0.157 $\big(\substack{0.236\\ 0.097}\big)$\\[1ex]
		
		& 3.&  neoplasms: 0.053 $\big(\substack{0.058\\ 0.048}\big)$& external: 0.074 $\big(\substack{0.085\\ 0.064}\big)$& digestive: 0.133 $\big(\substack{0.241\\ 0.066}\big)$\\[1ex]
		\hline\rule[10pt]{0pt}{0pt}
		
		&1.& external: 0.651 $\big(\substack{0.665\\ 0.634}\big)$ & external: 0.544 $\big(\substack{0.577\\ 0.498}\big)$&  external: 0.426 $\big(\substack{0.476\\ 0.348}\big)$\\[1ex]
		
		10--19 years& 2.& neoplasms: 0.099 $\big(\substack{0.108\\ 0.090}\big)$& nervous: 0.117 $\big(\substack{0.146\\ 0.091}\big)$& nervous: 0.155 $\big(\substack{0.211\\ 0.104}\big)$\\[1ex]
		
		& 3.&  nervous: 0.079 $\big(\substack{0.088\\ 0.071}\big)$& neoplasms: 0.110 $\big(\substack{0.133\\ 0.089}\big)$& neoplasms: 0.113 $\big(\substack{0.148\\ 0.079}\big)$\\[1ex]\hline\rule[10pt]{0pt}{0pt}
		
		&1.& external: 0.732 $\big(\substack{0.742\\ 0.721}\big)$ &external: 0.701 $\big(\substack{0.723\\ 0.672}\big)$& external: 0.653 $\big(\substack{0.690\\ 0.600}\big)$\\[1ex]
		
		20--29 years& 2.& neoplasms: 0.072 $\big(\substack{0.077\\ 0.067}\big)$& neoplasms: 0.084 $\big(\substack{0.097\\ 0.072}\big)$& neoplasms: 0.094 $\big(\substack{0.118\\ 0.074}\big)$\\[1ex]
		
		& 3.&  circulatory: 0.054 $\big(\substack{0.058\\ 0.049}\big)$& circulatory: 0.072 $\big(\substack{0.088\\ 0.059}\big)$& circulatory: 0.093 $\big(\substack{0.123\\ 0.069}\big)$\\[1ex]\hline\rule[10pt]{0pt}{0pt}
		
		&1.& external: 0.590 $\big(\substack{0.600\\ 0.579}\big)$ &external: 0.648 $\big(\substack{0.670\\ 0.622}\big)$& external: 0.679 $\big(\substack{0.712\\ 0.637}\big)$\\[1ex]
		
		30--39 years& 2.& circulatory: 0.125 $\big(\substack{0.132\\ 0.120}\big)$& circulatory: 0.125 $\big(\substack{0.140\\ 0.112}\big)$& circulatory: 0.120 $\big(\substack{0.142\\ 0.100}\big)$\\[1ex]
		
		& 3.&  neoplasms: 0.119 $\big(\substack{0.124\\ 0.113}\big)$& neoplasms: 0.095 $\big(\substack{0.105\\ 0.085}\big)$& neoplasms: 0.074 $\big(\substack{0.087\\ 0.062}\big)$\\[1ex]\hline\rule[10pt]{0pt}{0pt}
		
		&1.& external: 0.307 $\big(\substack{0.315\\ 0.298}\big)$ &external: 0.359 $\big(\substack{0.379\\ 0.337}\big)$& external: 0.398 $\big(\substack{0.429\\ 0.361}\big)$\\[1ex]
		
		40--49 years& 2.& 		
  neoplasms: 0.256 $\big(\substack{0.261\\ 0.249}\big)$& neoplasms: 0.233 $\big(\substack{0.246\\ 0.219}\big)$& neoplasms: 0.204 $\big(\substack{0.224\\ 0.183}\big)$\\[1ex]
	
			& 3.& circulatory: 0.205 $\big(\substack{0.211\\ 0.199}\big)$& circulatory: 0.148 $\big(\substack{0.158\\ 0.137}\big)$& circulatory: 0.104 $\big(\substack{0.116\\ 0.092}\big)$\\[1ex]
\hline\rule[10pt]{0pt}{0pt}
		
			&1.& neoplasms: 0.428 $\big(\substack{0.434\\ 0.423}\big)$ &neoplasms: 0.437 $\big(\substack{0.450\\ 0.422}\big)$& neoplasms: 0.409 $\big(\substack{0.434\\ 0.379}\big)$\\[1ex]
		
		50--59 years& 2.& circulatory: 0.229 $\big(\substack{0.234\\ 0.225}\big)$& external: 0.138 $\big(\substack{0.149\\ 0.128}\big)$& external: 0.148 $\big(\substack{0.165\\ 0.130}\big)$\\[1ex]
		
		& 3.&  external: 0.117 $\big(\substack{0.121\\ 0.113}\big)$& circulatory: 0.135 $\big(\substack{0.142\\ 0.128}\big)$& infectious: 0.102 $\big(\substack{0.148\\ 0.067}\big)$\\[1ex]\hline\rule[10pt]{0pt}{0pt}
		
		&1.& neoplasms: 0.500 $\big(\substack{0.505\\ 0.495}\big)$ &neoplasms: 0.568 $\big(\substack{0.578\\ 0.556}\big)$& neoplasms: 0.583 $\big(\substack{0.601\\ 0.563}\big)$\\[1ex]
		
		60--69 years& 2.& circulatory: 0.233 $\big(\substack{0.238\\ 0.229}\big)$& circulatory: 0.118 $\big(\substack{0.124\\ 0.113}\big)$& endocrine: 0.088 $\big(\substack{0.101\\ 0.077}\big)$\\[1ex]
		
		& 3.&  respiratory: 0.061 $\big(\substack{0.064\\ 0.059}\big)$& endocrine: 0.067 $\big(\substack{0.073\\ 0.062}\big)$& nervous: 0.067 $\big(\substack{0.078\\ 0.056}\big)$\\[1ex]\hline\rule[10pt]{0pt}{0pt}
		
		&1.& neoplasms: 0.432 $\big(\substack{0.436\\ 0.428}\big)$ &neoplasms: 0.514 $\big(\substack{0.524\\ 0.502}\big)$& neoplasms: 0.538 $\big(\substack{0.556\\ 0.514}\big)$\\[1ex]
		
		70--79 years& 2.& circulatory: 0.270 $\big(\substack{0.274\\ 0.266}\big)$& circulatory: 0.133 $\big(\substack{0.139\\ 0.128}\big)$& endocrine: 0.109 $\big(\substack{0.122\\ 0.096}\big)$\\[1ex]
		
		& 3.&  respiratory: 0.092 $\big(\substack{0.095\\ 0.089}\big)$& endocrine: 0.078 $\big(\substack{0.084\\ 0.073}\big)$& nervous: 0.077 $\big(\substack{0.088\\ 0.066}\big)$\\[1ex]\hline\rule[10pt]{0pt}{0pt}
		
		&1.& circulatory: 0.376 $\big(\substack{0.381\\ 0.371}\big)$ &neoplasms: 0.297 $\big(\substack{0.306\\ 0.285}\big)$& neoplasms: 0.301 $\big(\substack{0.320\\ 0.274}\big)$\\[1ex]
		
		80+ years& 2.& neoplasms: 0.259 $\big(\substack{0.262\\ 0.256}\big)$& circulatory: 0.244 $\big(\substack{0.255\\ 0.234}\big)$& mental: 0.160 $\big(\substack{0.222\\ 0.115}\big)$\\[1ex]
		
		& 3.&  respiratory: 0.112 $\big(\substack{0.116\\ 0.108}\big)$& mental: 0.091 $\big(\substack{0.115\\ 0.072}\big)$& circulatory: 0.146 $\big(\substack{0.158\\ 0.133}\big)$\\[1ex]
\hline
		\end{tabular}}
	\end{center}
\end{table}
\begin{table}[!ht]
	\begin{center}\footnotesize{
		\caption{Leading death causes with weights estimated by MCMC sample average with 5 and 95 percent
		confidence intervals of MCMC posterior in brackets for \emph{females} of all age categories in 2011, 2031 and
2051.}
		\label{tab:LeadingCausesFem}
		\begin{tabular}{rr|rrr}
			\cline{3-5}\rule[-3pt]{0pt}{12pt}
			  && 2011 & 2031 & 2051  \\
			\hline\rule[10pt]{0pt}{0pt}
		&1.& not elsewhere: 0.685 $\big(\substack{0.696\\ 0.673}\big)$ &not elsewhere: 0.631 $\big(\substack{0.658\\ 0.594}\big)$& not elsewhere: 0.531 $\big(\substack{0.584\\ 0.448}\big)$\\[1ex]
		
		0--9 years& 2.& external: 0.090 $\big(\substack{0.097\\ 0.084}\big)$& circulatory: 0.086 $\big(\substack{0.118\\ 0.061}\big)$& circulatory: 0.184 $\big(\substack{0.278\\ 0.111}\big)$\\[1ex]
		
		& 3.&  neoplasms: 0.057 $\big(\substack{0.063\\ 0.052}\big)$& neoplasms: 0.070 $\big(\substack{0.086\\ 0.055}\big)$& neoplasms: 0.076 $\big(\substack{0.104\\ 0.066}\big)$\\[1ex]
		\hline\rule[10pt]{0pt}{0pt}
		
		&1.& external: 0.507 $\big(\substack{0.527\\ 0.483}\big)$ & external: 0.439 $\big(\substack{0.482\\ 0.383}\big)$&  external: 0.363 $\big(\substack{0.425\\ 0.269}\big)$\\[1ex]
		
		10--19 years& 2.& neoplasms: 0.152 $\big(\substack{0.168\\ 0.137}\big)$& neoplasms: 0.169 $\big(\substack{0.208\\ 0.132}\big)$& neoplasms: 0.175 $\big(\substack{0.236\\ 0.114}\big)$\\[1ex]
		
		& 3.&  not elsewhere: 0.084 $\big(\substack{0.096\\ 0.073}\big)$& nervous: 0.104 $\big(\substack{0.141\\ 0.073}\big)$& nervous: 0.125 $\big(\substack{0.193\\ 0.070}\big)$\\[1ex]\hline\rule[10pt]{0pt}{0pt}
		
		&1.& external: 0.508 $\big(\substack{0.523\\ 0.491}\big)$ &external: 0.454 $\big(\substack{0.489\\ 0.414}\big)$& external: 0.397 $\big(\substack{0.448\\ 0.334}\big)$\\[1ex]
		
		20--29 years& 2.& neoplasms: 0.167 $\big(\substack{0.179\\ 0.156}\big)$& neoplasms: 0.185 $\big(\substack{0.213\\ 0.157}\big)$& neoplasms: 0.196 $\big(\substack{0.242\\ 0.151}\big)$\\[1ex]
		
		& 3.&  circulatory: 0.081 $\big(\substack{0.090\\ 0.073}\big)$& circulatory: 0.102 $\big(\substack{0.127\\ 0.080}\big)$& circulatory: 0.122 $\big(\substack{0.169\\ 0.083}\big)$\\[1ex]\hline\rule[10pt]{0pt}{0pt}
		
		&1.& external: 0.337 $\big(\substack{0.350\\ 0.324}\big)$ &external: 0.366 $\big(\substack{0.395\\ 0.333}\big)$& external: 0.381 $\big(\substack{0.427\\ 0.328}\big)$\\[1ex]
		
		30--39 years& 2.& neoplasms: 0.297 $\big(\substack{0.308\\ 0.286}\big)$& neoplasms: 0.233 $\big(\substack{0.254\\ 0.212}\big)$& neoplasms: 0.180 $\big(\substack{0.206\\ 0.152}\big)$\\[1ex]
		
		& 3.&circulatory: 0.120 $\big(\substack{0.128\\ 0.112}\big)$& circulatory: 0.124 $\big(\substack{0.143\\ 0.107}\big)$& circulatory: 0.123 $\big(\substack{0.153\\ 0.097}\big)$  \\[1ex]\hline\rule[10pt]{0pt}{0pt}
		
		&1.&		
  neoplasms: 0.480 $\big(\substack{0.488\\ 0.472}\big)$& neoplasms: 0.411 $\big(\substack{0.430\\ 0.390}\big)$& neoplasms: 0.338 $\big(\substack{0.365\\ 0.303}\big)$ \\[1ex]
		
		40--49 years& 2.& external: 0.164 $\big(\substack{0.171\\ 0.157}\big)$ &external: 0.189 $\big(\substack{0.207\\ 0.169}\big)$& external: 0.204 $\big(\substack{0.235\\ 0.170}\big)$\\[1ex]
	
			& 3.& circulatory: 0.133 $\big(\substack{0.138\\ 0.127}\big)$& circulatory: 0.113 $\big(\substack{0.124\\ 0.102}\big)$& circulatory: 0.093 $\big(\substack{0.108\\ 0.078}\big)$\\[1ex]
\hline\rule[10pt]{0pt}{0pt}
		
			&1.& neoplasms: 0.588 $\big(\substack{0.594\\ 0.581}\big)$ &neoplasms: 0.586 $\big(\substack{0.601\\ 0.569}\big)$& neoplasms: 0.553 $\big(\substack{0.578\\ 0.520}\big)$\\[1ex]
		
		50--59 years& 2.& circulatory: 0.120 $\big(\substack{0.124\\ 0.116}\big)$& external: 0.092 $\big(\substack{0.102\\ 0.082}\big)$& external: 0.105 $\big(\substack{0.122\\ 0.088}\big)$\\[1ex]
		
		& 3.&  external: 0.076 $\big(\substack{0.079\\ 0.072}\big)$& circulatory: 0.063 $\big(\substack{0.067\\ 0.058}\big)$& not elsewhere: 0.063 $\big(\substack{0.082\\ 0.050}\big)$\\[1ex]\hline\rule[10pt]{0pt}{0pt}
		
		&1.& neoplasms: 0.566 $\big(\substack{0.571\\ 0.561}\big)$ &neoplasms: 0.640 $\big(\substack{0.651\\ 0.626}\big)$& neoplasms: 0.666 $\big(\substack{0.684\\ 0.640}\big)$\\[1ex]
		
		60--69 years& 2.& circulatory: 0.157 $\big(\substack{0.161\\ 0.153}\big)$& respiratory: 0.075 $\big(\substack{0.083\\ 0.069}\big)$& nervous: 0.067 $\big(\substack{0.084\\ 0.058}\big)$\\[1ex]
		
		& 3.&  respiratory: 0.079 $\big(\substack{0.083\\ 0.076}\big)$& circulatory: 0.066 $\big(\substack{0.069\\ 0.062}\big)$& respiratory: 0.067 $\big(\substack{0.077\\ 0.057}\big)$\\[1ex]\hline\rule[10pt]{0pt}{0pt}
		
		&1.& neoplasms: 0.405 $\big(\substack{0.409\\ 0.400}\big)$ &neoplasms: 0.478 $\big(\substack{0.490\\ 0.464}\big)$& neoplasms: 0.495 $\big(\substack{0.517\\ 0.465}\big)$\\[1ex]
		
		70--79 years& 2.& circulatory: 0.256 $\big(\substack{0.260\\ 0.252}\big)$& respiratory: 0.114 $\big(\substack{0.124\\ 0.106}\big)$& respiratory: 0.115 $\big(\substack{0.130\\ 0.101}\big)$\\[1ex]
		
		& 3.&  respiratory: 0.100 $\big(\substack{0.104\\ 0.097}\big)$& circulatory: 0.111 $\big(\substack{0.116\\ 0.106}\big)$& nervous: 0.094 $\big(\substack{0.108\\ 0.081}\big)$\\[1ex]\hline\rule[10pt]{0pt}{0pt}
		
		&1.& circulatory: 0.430 $\big(\substack{0.435\\ 0.424}\big)$ &circulatory: 0.252 $\big(\substack{0.264\\ 0.239}\big)$& mental: 0.255 $\big(\substack{0.338\\ 0.189}\big)$\\[1ex]
		
		80+ years& 2.& neoplasms: 0.167 $\big(\substack{0.169\\ 0.164}\big)$& neoplasms: 0.180 $\big(\substack{0.188\\ 0.170}\big)$& neoplasms: 0.163 $\big(\substack{0.178\\ 0.143}\big)$\\[1ex]
		
		& 3.&  respiratory: 0.093 $\big(\substack{0.097\\ 0.090}\big)$& mental: 0.151 $\big(\substack{0.187\\ 0.121}\big)$& circulatory: 0.130 $\big(\substack{0.144\\ 0.115}\big)$\\[1ex]
\hline
		\end{tabular}}
	\end{center}
\end{table}

\bibliography{pension}
\bibliographystyle{chicago} 

\end{document}